\documentclass[preprint,prd,nofootinbib,showpacs,showkeys]{revtex4}
\usepackage[dvips]{epsfig}
\usepackage{epsf,amsfonts,amssymb}
\usepackage{hyperref}

\newcommand{\gs}{\ensuremath{g_s}}      
\newcommand{\alp}{\ensuremath{\alpha'}} 

\newcommand{\eps}{\ensuremath{\epsilon}}
\newcommand{\z}{\ensuremath{\zeta}}

\newcommand{\tr}{\mathop{\rm Tr}}
\newcommand{\half}{\ensuremath{\frac{1}{2}}}
\newcommand{\bbR}[1]{\mbox{${\mathbb R}^{#1}$}}
\newcommand{\bbZ}[1]{\mbox{${\mathbb Z}_{#1}$}}
\newcommand{\w}{\wedge}

\newcommand{\cA}{\ensuremath{\mathcal{A}}}
\newcommand{\cN}{\ensuremath{\mathcal{N}}}
\newcommand{\cO}{\ensuremath{\mathcal{O}}}
\newcommand{\cT}{\ensuremath{\mathcal{T}}}

\newcommand{\eg}{\textit{e.g}}
\newcommand{\ie}{\textit{i.e}}

\newcommand{\N}[1]{\ensuremath{\cN=#1}}
\newcommand{\F}[1]{\ensuremath{F_{[#1]}}}
\newcommand{\ads}[1]{\mbox{${AdS}_{#1}$}}
\newcommand{\adss}[2]{\mbox{$AdS_{#1}\times {S}^{#2}$}}

\newcommand{\Gc}{\ensuremath{\Gamma_{\mathrm{cusp}}}}        
\newcommand{\bGc}{\ensuremath{\bar{\Gamma}_{\mathrm{cusp}}}} 
\newcommand{\tc}{\ensuremath{g_sN}}                          
\newcommand{\gN}{\ensuremath{\sqrt{g_sN}}}                   
\newcommand{\gym}{\ensuremath{g_{\mathrm{YM}}}}              
\newcommand{\Too}{\ensuremath{{\mathbb T}^{1,1}}}            
\newcommand{\pD}{\ensuremath{p.\Delta}}
\newcommand{\Tau}{\ensuremath{\cT}}                          
\newcommand{\trz}{\ensuremath{\tilde{r}_0}}
\newcommand{\trho}{\ensuremath{\tilde{\rho}}}

\begin{document}

\title{ Wilson loops and anomalous dimensions in cascading theories}

\date{October 2003}

\author{Martin Kruczenski}

\affiliation{Department of Physics, Brandeis University\\
             Waltham, MA 02454.}

\email[e-mail:]{martink@brandeis.edu}

\begin{abstract}
We use light-like Wilson loops and the AdS/CFT correspondence to 
compute the anomalous dimensions of twist two operators in the
cascading (Klebanov-Strassler) theory. The computation amounts
to find a minimal surface in the UV region of the KS background
which is described by the Klebanov-Tseytlin solution.  
The result is similar to the one for $SU(N)$ \N{4} SYM but with $N$ 
replaced by an effective, scale dependent $N$. We performed also a
calculation using a rotating string and find agreement. In fact we use 
a double Wick rotated version of the rotating string solution which is an Euclidean
world-sheet ending on a light-like line in the boundary. It gives the same result 
as the rotating string for the \N{4} case but is more appropriate than the rotating 
string in the \N{1} case. 
\end{abstract}

\preprint{BRX TH-256}
\preprint{hep-th/0310030}

\maketitle

\section{Introduction}

 String theory originated as an attempt to understand strong interactions and in particular
the hadronic spectrum. Later it was superseded by QCD which successfully explained 
the experimental data on deeply inelastic scattering (DIS) and other processes involving 
large energy scales where, due to asymptotic freedom, the coupling constant was small. 
In spite of that success, the problem of computing the hadronic
spectrum in QCD is still unsolved. To study this problem 't Hooft proposed 
to consider gauge theories in the limit of large number of colors \cite{largeN1,largeN2} 
a limit in which the gauge theory actually seems to behave as a theory of strings. Interestingly, 
for certain 4-dimensional gauge theories, this idea was recently put in a remarkably precise form by Maldacena 
through the AdS/CFT correspondence\cite{malda}. 
 In its most studied example, the conjectured AdS/CFT correspondence 
\cite{malda,Gubser:1998bc,Witten:1998qj} provides a description of the large N limit 
of \N{4} SU(N) SYM theory as type IIB string theory on $AdS_5\times S^5$. Starting 
from there, much understanding has been gained on the behavior 
of other SYM theories in the 't Hooft limit\footnote{There is a large literature on the subject. 
See \cite{magoo} for a comprehensive review}. 
In particular in this paper we 
consider the \N{1} background which was derived in \cite{Klebanov:2000nc} and is dual to a cascading 
theory. This field theory is somewhat peculiar and we describe its properties in 
section \ref{sec:KTbkg}.

 Returning for a moment to QCD, in the comparison of theory with experiments an important role 
is played by twist two operators which are operators with the lowest conformal dimension 
(in the free theory) for a given spin. They give the leading contribution at large momentum 
transfer for scattering of leptons by hadrons (DIS)\cite{DIS1,DIS2}. Their anomalous dimensions determine 
the corrections to Bjorken scaling of the cross section which is a fundamental point in comparison 
between theory and experiment.

 It is natural then to wonder if these anomalous dimensions can be computed using AdS/CFT in the strongly
coupled regime. Unfortunately there is still no known dual to QCD but in the case of \N{4} SYM 
such calculation can be done. In the \N{4} theory one can construct the following operators:
\begin{eqnarray}
\cO^\Phi_{(\mu_1\cdots\mu_S)} &=& \tr \Phi^I \nabla_{(\mu_1} \cdots \nabla_{\mu_S)} \Phi^I, \nonumber\\ 
\cO^\Psi_{(\mu_1\cdots\mu_S)} &=& \tr \Psi^a \gamma_{(\mu_1} \nabla_{\mu_2} \cdots \nabla_{\mu_S)} \Psi^a, \\
\cO^F_{(\mu_1\cdots\mu_S)} &=& \tr F^{\alpha}_{(\mu_1} \nabla_{\mu_2} \cdots \nabla_{\mu_{S-1}} F_{\mu_S)\alpha}, \nonumber 
\label{eq:t2op}
\end{eqnarray}
where $(\ldots )$ indicates traceless symmetrization. In the free theory, they have
the property of being the spin $S$ operators with smallest conformal dimension $\Delta=S+2$. 
In the limit $N\rightarrow\infty$, with $\tc$ fixed and large the corresponding calculation 
was done in \cite{Gubser:2002tv}. What they found in that paper was that, in the large 't Hooft coupling regime,
the operator with smallest conformal dimension for a given spin is
dual (in the AdS/CFT sense) to a rotating string in global $\adss{5}{5}$.
It was not clear that such operators were the same as those with the same property at
small coupling, namely (\ref{eq:t2op}). It was suggestive nevertheless that at strong coupling, the 
anomalous dimensions behaved as 
\begin{equation}
\Delta \simeq S + \frac{\gN}{\pi} \ln S, \ \ \ \ (S\rightarrow\infty)
\label{eq:andim}
\end{equation}
for large spin $S$. The point is that a similar logarithmic behavior is found
in perturbation theory at 1-loop \cite{DIS1,DIS2,Gubser:2002tv} and 2-loops \cite{Axenides:2002zf,Kotikov:2003fb,Kotikov:2002ab} for 
operators of the type (\ref{eq:t2op}) suggesting an identification between them and the rotating strings. 

In \cite{llWl} and \cite{Makeenko:2002qe} this idea was confirmed by using known field theory arguments  
to relate the computation of the anomalous dimension of twist two operators to 
the evaluation of a certain light-like Wilson loop. This Wilson loop can be evaluated 
using AdS/CFT and gives the same result (\ref{eq:andim}). This alternative method is applied directly in Poincare 
coordinates where the boundary is \bbR{(3,1)}. This is useful since many backgrounds are
known, or become simpler, in these coordinates. It also makes evident the use of symmetries in the
problem. For example the logarithmic dependence in the spin $S$ follows automatically from Lorentz
invariance. Moreover, in the case of \adss{5}{5}, the surface that computes the relevant Wilson  
loop can be found using only symmetry arguments. This type of calculation was extended to other cases 
in \cite{Belitsky:2003ys}.

 As a comment, let us emphasize that even if both methods use the AdS/CFT correspondence, they do 
so in a very different way and that such agreement should exist is not at all obvious from supergravity. 
 This agreement also shows that, although the identification of rotating strings and twist two operators
goes beyond the usual identification between supergravity modes and chiral primary operators (in the 
spirit of\cite{Berenstein:2002jq}), in a certain sense that identification is contained in the
identification between Wilson loops and string world-sheets. Clearly this last 
identification \cite{Maldacena:1998im,Rey:1998ik} also goes beyond supergravity since it makes explicit use of string world-sheets.  

 In view of the interest of twist two operators in QCD, it is important to repeat the \N{4} calculation in 
other, less supersymmetric backgrounds. The anomalous dimension is an ultraviolet  
property of the theory so we have to study backgrounds which are not asymptotically $\ads{5}$
since an asymptotically $\ads{5}$ background will give the same result as \ads{5}. One such background 
is the Klebanov-Strassler \cite{Klebanov:2000hb} that we mentioned before.  In section \ref{sec:Wl} we 
compute the anomalous dimension of twist two operators in that background.

After that, we also perform the rotating string calculation and compare the leading order results finding agreement 
between the two methods. In order to do that we need to use a different coordinate system and a different solution from the
one used in \cite{Gubser:2002tv}. The solution we use is an Euclidean world-sheet ending on a light-like line. 
The reason being that in that case the relevant part of the world-sheet is completely contained in the UV region
of the background and we can use the Klebanov-Tseytlin solution. This is reasonable since the anomalous dimension is a UV
property of the theory. As we shall see using this world-sheet also seems appropriate since the twist two operators 
appear in the Taylor expansion of a bilocal operator containing a light-like Wilson line which is precisely where the 
world-sheet ends.


\section{Anomalous dimensions of twist two operators from supergravity}
\label{sec:rotstringWl}

 The calculation of the anomalous dimensions of twist two operators can be done in supergravity
using rotating strings \cite{Gubser:2002tv} or Wilson loops \cite{llWl,Makeenko:2002qe}. Since the 
method using Wilson loops is formulated in Poincare coordinates, it can be directly applied in this case. 
Nevertheless we shall see presently that the rotating string method, after a small modification of the original calculation,  
can also be applied in Poincare coordinates 

 In this section we review both methods for the case \N{4} SYM.

\subsection{\label{adimfrt} Anomalous dimensions from rotating strings}

Following \cite{Gubser:2002tv}, to compute the anomalous dimension of an operator using rotating strings,
we should use a state operator map. Such a map consists in first doing a Wick rotation from $\bbR{(3,1)}$
into Euclidean space and then a conformal transformation from $\bbR{4}$ to $S^3\times\bbR{1}$: 
\begin{equation}
ds^2 = \sum_{i=1}^{4} dx_i^2 
     = e^{2\tau} \left(d\tau^2 + d\Omega_3^2\right)\ \ 
 \Longrightarrow \ \ ds'^2 = d\tau^2 + d\Omega_3^2 ,
\end{equation}
where $e^{2\tau} = \sum_{i=1..4} x_i^2$. Finally we do a Wick rotation on the $\bbR{1}$ to 
get a theory on $S^3\times \bbR{1}$ in Lorentzian signature. If the theory is conformally 
invariant\footnote{If it is not conformally invariant it is better to use other coordinates such that the conformal 
dimension is directly related to translations in a certain coordinate with no need of a Wick rotation as we describe below.}
we obtain the same theory on $S^3\times \bbR{1}$ and the conformal dimension of an
operator in $\bbR{(3,1)}$ maps into the energy of a corresponding state in $S^3\times \bbR{1}$. 
The problem of finding the operator of lowest conformal dimension
for a given spin becomes the problem of finding the state of lowest energy for a given spin.
In the limit $N\rightarrow\infty$, $g_s\rightarrow 0$ with $g_sN$ finite and large we can use 
the AdS/CFT correspondence\cite{malda,Gubser:1998bc,Witten:1998qj}. \N{4} SYM in this limit 
is dual to $\adss{5}{5}$ in global coordinates. The metric is:
\begin{equation}
ds^2 = R^2 \left(-\cosh^2\!\rho\,dt^2 + d\rho^2 + \sinh^2\!\rho\,d\Omega_3^2 + d\Omega_5^2 \right).
\end{equation} 
The state of minimal energy for a given spin $S$ turns out to be a string with its center of mass
fixed at $\rho=0$ and rotating with spin $S$. For large spin the energy of that string 
is (at leading order) \cite{Gubser:2002tv}:
\begin{equation}
E \simeq S + \frac{\gN}{\pi} \ln S , \ \ \ \ \ (S\rightarrow\infty) ,
\label{rotstringN4}
\end{equation}
where $\gs=\gym^2$. After identifying $E=\Delta$ this formula becomes eq.(\ref{eq:andim}).

 For the purpose of the following sections it is useful to repeat the calculation in a slightly
different way. The idea is to do the double Wick rotation directly in the solution. We obtain
then an Euclidean world-sheet completely contained in the Poincare patch. This is actually equivalent to using 
a system of coordinates such that scale transformations and boosts correspond to translations in certain coordinates.
In that way, the conserved momenta of the solution are directly related to the conformal dimension and
eigenvalue under a boost, without the need of any Wick rotations. 

This coordinate system, as well as other related coordinates are described in detail 
in Appendix A. Here we just state its relation to Poincare coordinates:
\begin{equation}
\begin{array}{lclllcl}
x_1 &=& R\, e^{-\tau}\,\tanh\rho\,  \sin\phi \cos\theta &,&
x_2 &=& R\, e^{-\tau}\,\tanh\rho\,  \sin\phi \sin\theta , \\
x_3 &=& R\, e^{-\tau}\,\tanh\rho\,  \cos\phi \cosh\chi &,&
t   &=& R\, e^{-\tau}\,\tanh\rho\,  \cos\phi \sinh\chi  , \\
z &=& R\, e^{-\tau}\, \frac{1}{\cosh(\rho)} . &\ \  &&& 
\end{array}
\label{Poinnew}
\end{equation}
 From here we see first that $z>0$, \ie this patch is embedded in the Poincare patch, and second that
$|x_3|>|t|$ implying that only that region is covered which will turn out to be enough for our purposes. 
The \ads{5} metric in this new coordinates is
\begin{equation}
ds^2 = R^2 (\cosh^2\!\rho\,d\tau^2 + d\rho^2 + \sinh^2\!\rho\,d\phi^2 + \sinh^2\!\rho\,\sin^2\!\phi\,d\theta^2
            -\sinh^2\!\rho\,\cos^2\!\phi\,d\chi^2) ,
\label{N4metric}
\end{equation}
which has as explicit isometries translations in $\tau$, $\chi$ and $\theta$ giving rise to three conserved momenta
$(P_{\tau},P_{\chi},P_{\theta})$. Translations in $\tau$ correspond to
scale transformations $z\rightarrow \lambda z$, $x_i\rightarrow \lambda x_i$ and $t\rightarrow\lambda t$ and then 
we should identify $P_{\tau}$ with the conformal dimension $\Delta=P_\tau$. Translations in 
$\chi$ and $\theta$ correspond to boosts and rotations respectively.

As discussed before, in global coordinates $\Delta$ was identified with the energy but that depended
on a double Wick rotation. Here no Wick rotation is necessary since the generator of scale transformations and $\tau$ 
translations are the same. 

The `rotating' string solution is now an Euclidean world-sheet given by
\begin{equation}
\begin{array}{l}
\rho=\sigma, \ \ \ \tau=\tau, \ \ \ \chi=\omega\tau, \ \ \ \phi=\theta=0 , \nonumber\\
-\rho_0<\rho<\rho_0, \ \ \ \coth(\rho_0) = \omega ,
\end{array}
\label{rotansatzN4}
\end{equation}
which determines the momenta to be:
\begin{equation}
\begin{array}{lcccl}
P_{\tau} &=& \Delta &=& 4\frac{R^2}{2\pi\alpha'} 
                        \int_0^{\sigma_0} d\sigma \frac{\cosh^2\sigma}{\sqrt{\cosh^2\sigma-\omega^2\sinh^2\sigma}} ,\\
P_{\chi} &=& S &=&  4\frac{R^2}{2\pi\alpha'} 
                    \int_0^{\sigma_0} d\sigma \frac{\omega\sinh^2\sigma}{\sqrt{\cosh^2\sigma-\omega^2\sinh^2\sigma}} ,\\
P_{\theta} &=& 0 , &&
\end{array}
\end{equation}
where we used the world-sheet Euclidean coordinate $\tau$ as (world-sheet) time to define the conserved charges. Expanding this 
integrals for large spin $S$ we obtain the result:
\begin{equation}
\Delta \simeq S + \frac{R^2}{\pi\alp} \ln S ,
\label{N4rotstring}
\end{equation}
which is the same as  (\ref{rotstringN4}) after using $R^2/\alp=\sqrt{g_sN}$.

\begin{figure}
\centerline{\epsfysize=08cm\epsfbox{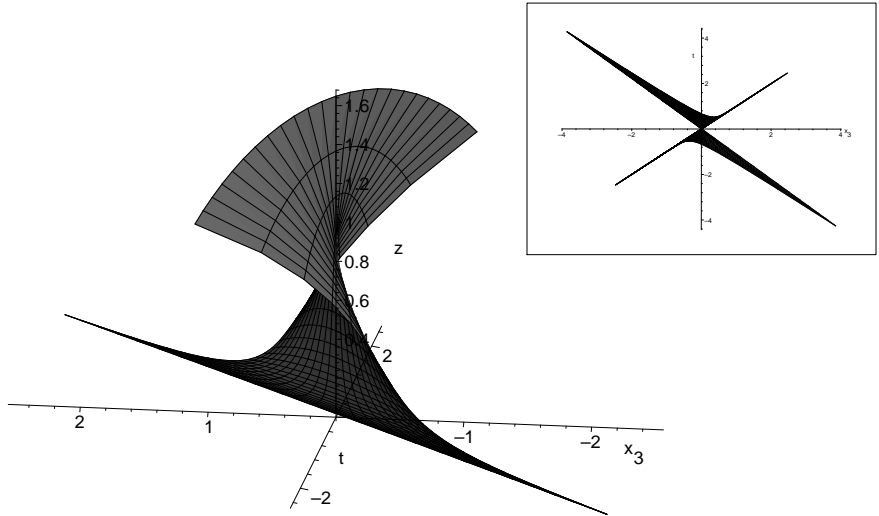}}
\caption{\small `Rotating` string in \ads{5} space in Poincare coordinates. This is an Euclidean 
world-sheet which approaches the boundary at the $x_3=t$ line. The results for the anomalous dimension 
are the same as with the usual rotating string in global coordinates.
The inset depicts the projection of the world-sheet onto the boundary on the plane ($x_3$,$t$).}
\label{fig:rotstring}
\end{figure}

In Poincare coordinates the solution (\ref{rotansatzN4}) reads
\begin{equation}
\begin{array}{lclclcl}
x_3&=&e^{-\tau}\, \tanh\sigma\,  \cosh(\omega\tau),&\ \ \ \ & t &=& e^{-\tau}\, \tanh\sigma\,  \sinh(\omega\tau) , \\
z&=& e^{-\tau}\, \frac{1}{\cosh\sigma} , && x_{1,2}&=&0 .
\end{array}
\end{equation}
This world-sheet is depicted in figure \ref{fig:rotstring}. Notice that as $\tau\rightarrow\infty$ the string
extends along $x_3=t$ at a small value of $z$, namely in the UV of the field theory. This is in agreement
with the field theory expectations since the anomalous dimension is a UV quantity, the usual rotating string calculation
seems to mix UV and IR. This is not important for the conformal case but is relevant for the non-conformal one we
discuss later on.

As an aside note also that as $\tau$ goes to infinity which is the relevant limit to compute the anomalous dimension, the
string extends, as we just said, along the light-like line $x_3=t$. It will be interesting to understand if this
is in some way related to the Wilson loop calculation which also uses light-like lines in the boundary. On the other 
hand it seems to be related to the fact that the twist two operators appear in the expansion of a bilocal operator 
which contains a light-like Wilson line as we see from eq.(\ref{eq:OsD}) below.

 \subsection{\label{adimfWl} Anomalous dimensions from Wilson loops}

\begin{figure}
\centerline{\epsfysize=10cm\epsfbox{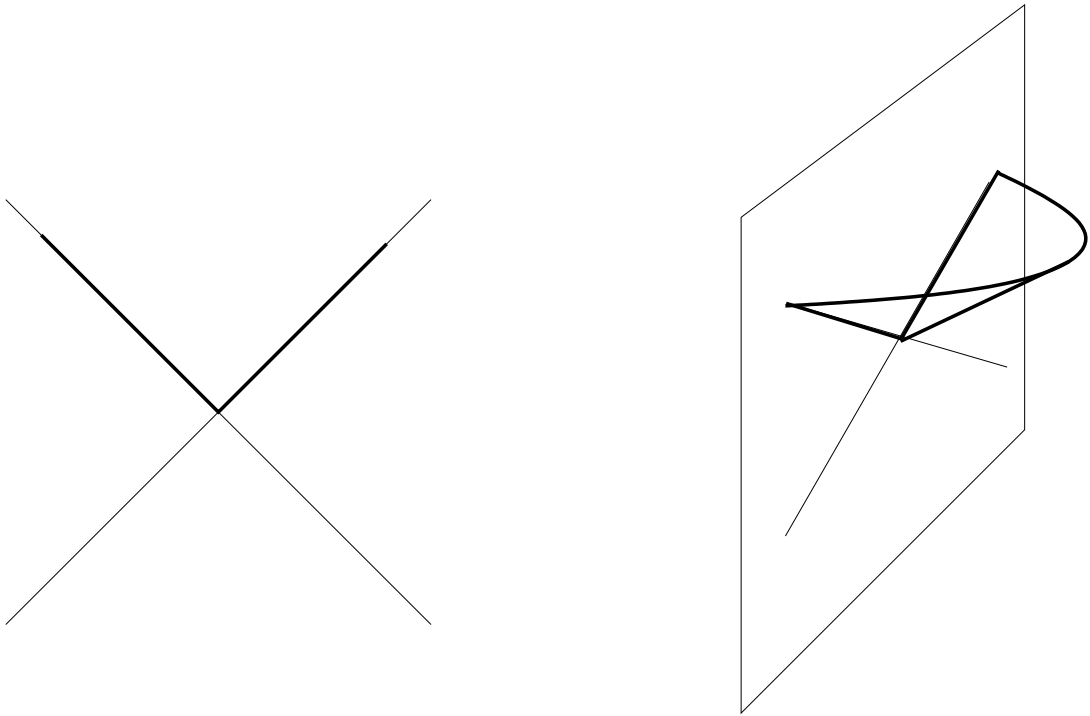}}
\caption{\small Light-like Wilson loop used to compute the anomalous dimension and the corresponding surface in \ads{5}. Thin
lines indicate the light-cone. This Wilson loop is actually singular but can be used to compute the limit of a space-like one
since the \ads{5} surface is well defined. }
\label{fig:cuspN4}
\end{figure}

 In gauge field theories there is a way to compute the anomalous dimensions of large spin, twist
two operators using Wilson loops \cite{Korchemsky,Korchemsky:1989si}. The argument starts by considering the bilocal 
operator:
\begin{eqnarray}
W_{\Delta^\mu} &=& \tr \left(\Phi^I(\Delta^\mu) e^{\int_0^{1} A_\mu(t\Delta^\mu)\Delta^\mu dt} \Phi^I(0) \right)
 = \sum_{S=0}^{\infty} \frac{1}{S!}\cO_S(\Delta^\mu), \\
\cO_S(\Delta^\mu) &=& \tr \left(\Phi^I \nabla_{\mu_1} \cdots \nabla_{\mu_S} \Phi^I \right) \Delta^{\mu_1}\cdots \Delta^{\mu_S},
\label{eq:OsD}
\end{eqnarray}
where $\Delta^{\mu}$ is a fixed light-like vector and the Wilson line is in the adjoint representation 
($A_\mu=A^a_\mu C_{abc}$). The operators $\cO_S(\Delta^\mu)$ are the (space-time) component of the operators
(\ref{eq:t2op}) with maximum helicity along $\Delta^\mu$.
 Matrix elements of this operator require renormalization which give rise to an anomalous dimension.
In particular we can compute matrix elements between one particle states of the field $\Phi^I$ with momentum $p$.
 It was observed in \cite{Korchemsky,Korchemsky:1989si} that the limit $S\rightarrow\infty$ corresponds to $p.\Delta\rightarrow\infty$
and that, in that limit, such matrix elements can be computed using Wilson loops. The relevant Wilson loops have 
an anomalous dimension due to the fact that they have cusps. Generically, for a Wilson loop with a cusp of angle
$\gamma$ the expectation value behaves as \cite{Polyakov}:
\begin{equation} 
W_{\gamma} \sim \left(\frac{\Lambda}{m}\right)^{-\Gc(\gamma)} ,
\label{eq:cuspW}
\end{equation}
where $\Lambda$ is an UV cutoff. In our case it turns out that $\gamma \simeq \ln(p.\Delta)$ which diverges
for $\pD\rightarrow\infty$. In such regime the cusp anomaly behaves as \cite{gcusp}:
\begin{equation}
\Gc(\gamma)\simeq \bGc \gamma, \ \ \ \gamma\rightarrow\infty .
\label{eq:Gdiv}
\end{equation}
Using this result one can see that the anomalous dimension is given by:
\begin{equation}
\Delta_S - S -2 = \gamma_S \simeq -2\bGc \ln S,\ \ \ \ (S\rightarrow\infty).
\label{eq:adfWl}
\end{equation}
 What is left now is to compute $\bGc$. To summarize, one should compute a Wilson loop with the shape of a wedge 
such as that in fig.\ref{fig:cuspN4} where both sides approach the light-like lines $x=\pm t$, $t>0$. In that limit
the expectation value should behave as in (\ref{eq:cuspW}) with $\Gc$ as in eq.(\ref{eq:Gdiv}). After extracting the 
constant $\bGc$, we can use (\ref{eq:adfWl}) and get the anomalous dimension. A Wilson loop as that in fig.\ref{fig:cuspN4}
can be computed by using a surface of minimal area in \ads{5} with metric\footnote{Wilson loops with cusps where first computed 
using AdS/CFT in Euclidean signature in \cite{Drukker:1999zq}.}
\begin{equation}
ds^2 = \frac{R^2}{z^2} \left( dz^2 -dt^2 + dx_1^2 +dx_2^2 +dx_3^2 \right) ,
\end{equation}
and ending at $z=0$ on the lines $x=\pm t$, $t>0$. 
Such surface turns out to be given simply by the equation $z^2=2(t^2-x^2)$. The area \cA\ of such surface is 
divergent but we can regulate it as:
\begin{equation}
\cA = \frac{R^2}{2\pi\alpha'}\frac{1}{2} \int_0^{\infty} \frac{d\rho}{\rho} \int_{-\infty}^{+\infty} d\xi
  = \frac{R^2}{2\pi\alpha'}\frac{1}{2} \int_\eps^{L} \frac{d\rho}{\rho} \int_{-\gamma/2}^{+\gamma/2} d\xi 
  = \frac{R^2}{2\pi\alpha'}\frac{1}{2} \gamma \ln \frac{L}{\epsilon} ,
\label{eq:N4area}
\end{equation}
which, using that $\cA=\ln W$, $\Lambda\sim 1/\epsilon$ and $R^2/\alp=\sqrt{\tc}$  implies that $\bGc=-\gN /4\pi$. 
This calculation is actually for a Wilson loop in the fundamental. Since we
need a Wilson loop in the adjoint we should multiply the result by a factor of two. Taking this into account 
and using formula (\ref{eq:adfWl}) we obtain the conformal dimension
\begin{equation}
\Delta_S \simeq S + \frac{\gN}{\pi} \ln S, \ \ \ \ \ (S\rightarrow\infty) .
\end{equation} 
Actually, the surface $z^2=2(t^2-x^2)$ has a simple expression in embedding coordinates which
can be obtained from symmetry arguments alone. A comment is that one can be more rigorous and find
a surface ending on two space-like lines and take the limit in which they become light-like obtaining
the same result. 

The fact that the anomalous dimension behaves as $\ln S$ for large $S$ depends on the fact that the cusp 
anomaly $\Gc(\gamma)$ diverges linearly in $\gamma$ for large $\gamma$. In the Wilson loop calculation 
this follows from the fact that the integrand in eq.(\ref{eq:N4area}) does not depend on $\xi$ which is a consequence 
of Lorentz invariance. It is interesting to observe that the fact that the anomalous dimension behaves 
as $\ln S$ \cite{gcusp} appears as a consequence of Lorentz invariance also in the field theory 
calculation\footnote{I am grateful to G. Sterman for pointing out this fact to me.}.

It is also interesting to note that the symmetries determine the behavior of the cusp anomaly for small
Euclidean angles. In fact, the flat metric $ds^2 = dr^2 + r^2 d\theta^2$, 
is invariant (up to a conformal factor) under the transformation 
$r\rightarrow r^\lambda $, $\theta\rightarrow \lambda\theta$. When $\Theta\rightarrow 0$ we can ignore the
fact that such a transformation alters the periodicity of $\theta$ and obtain that $\ln W_M \sim - \Gc(\Theta) \ln (L/\eps)$
is invariant only if $\Gc(\Theta) \sim 1/\Theta (\Theta\rightarrow 0)$. From supergravity we get   
\begin{equation}
\Gamma_{\mathrm{cusp}}(\Theta) \simeq  - \frac{R^2}{2\pi\alp} \frac{c^2}{\Theta}, \ \ \ \ (\Theta\rightarrow 0).
\label{eq:Theta0}
\end{equation}
with the expected behavior $1/\Theta$. The constant $c$ is given by \cite{llWl}:
\begin{equation}
c = \int_{-\infty}^{\infty} \frac{du}{(1+u^2)^{3/2}(2+u^2)^{1/2}} 
  = \frac{(2\pi)^{\frac{3}{2}}}{\left(\Gamma(\frac{1}{4})\right)^{2}}.
\label{eq:c}
\end{equation}
 We can also compute the same result\footnote{This was pointed out to me by J. Maldacena.}
 by using the quark anti-quark potential $V(L)$ which was computed in \cite{Maldacena:1998im,Rey:1998ik}. 
In fact, when $\Theta\rightarrow 0$ the lines are almost parallel and the Wilson loop should be
\begin{equation}
 \ln W \simeq \int_{\epsilon}^{L} d\rho V(\rho) =  - \frac{2c_1^2}{\pi} \gN \int_{\epsilon}^{L} \frac{d\rho}{\rho\Theta} ,
\end{equation}
and therefore the cusp anomaly is
\begin{equation}
\Gc \simeq -\frac{2c_1^2}{\pi} \gN \frac{1}{\Theta} ,
\end{equation}
which agrees with eq.(\ref{eq:Theta0}) if we use (\ref{eq:c}) and the value $c_1=\sqrt{2}\pi^{3/2}/\Gamma(1/4)^2$ 
from \cite{Maldacena:1998im,Rey:1998ik}.


\section{The Klebanov-Tseytlin background and Cascading theories.}
\label{sec:KTbkg}

 In this section we review the Klebanov-Tseytlin background and its 
field theory dual. 

\subsection{\label{sugrabkg} Supergravity background}

In \cite{Klebanov:2000nc}, I. Klebanov and A. Tseytlin found a type $IIB$, \N{1} background with metric
\begin{equation}
ds^2=h(r)^{-\half} dx_{\mu}dx^{\mu} + h(r)^{\half} ds^2_{[6]} 
    =h(r)^{-\half} dx_{\mu}dx^{\mu} + h(r)^{\half} \left(dr^2+r^2 ds^2_{\Too} \right) ,
\label{eq:KTmetric}
\end{equation}
where $ds^2_{[6]} = dr^2+r^2 ds^2_{\Too} $ is a Ricci-flat K\"ahler metric on the complex cone
$w_1^2+w_2^2+w_3^2+w_4^2=0$, $\Too$ is a manifold which is the base of the cone and 
we defined
\begin{eqnarray}
h(r) &=& \frac{1}{r^4}\left[R^4+2L^4\left(\ln\frac{r}{r_0}+\frac{1}{4}\right)\right] , \label{eq:h(r)}\\
R^4 &=& \frac{27}{4} \gs N \pi (\alp)^2 , \\
L^2 &=& \frac{9}{4} \gs M \alp . \label{eq:Ldef}
\end{eqnarray}
 Here $M$ and $N$ are arbitrary parameters of the solution. The RR and NSNS field strengths are given by:
\begin{eqnarray}
\F{5} &=& (1+*_{10}) (\partial_r h^{-1}) d^4x\w dr  , \\
B_{[2]} &=& \frac{2L^2}{3} \ln(r/r_0) \omega_{[2]} , \\
\F{3} &=& \frac{2L^2}{9\gs}\omega_{[3]} ,
\end{eqnarray}
where $\omega_{[3]}$ and $\omega_{[2]}$ are certain closed forms in $\Too$ which can be found in \cite{Klebanov:2000nc,Klebanov:2000hb}
and will not be essential to our purpose here. 

The complex cone $w_1^2+w_2^2+w_3^2+w_4^2=0$ can be endowed with a K\"ahler metric whose K\"ahler potential 
depends only on $\sum_{i=1}^4 w_i\bar{w}_i$ which is actually the metric $ds^2_{[6]}$ we are using. 
It follows that written in terms of the $w_i$ the K\"ahler potential has an $SU(4)$ symmetry which
however is reduced to $SO(4)\times U(1)$ due to the condition $w_1^2+w_2^2+w_3^2+w_4^2=0$. The $SO(4)$ acts on the indices $i$
and the $U(1)$ symmetry is $w_i\rightarrow e^{i\alpha} w_i$. This $SO(4)\times U(1)$ is therefore an isometry of the 
metric (\ref{eq:KTmetric}). The $U(1)$ however is not a symmetry \cite{Klebanov:2002gr} because, in spite of being an 
isometry, it does not leave invariant the RR potential $C_{[2]}$ ($\F{3}=dC_{[2]}$). This corresponds in the field theory 
to a chiral anomaly that breaks the $U(1)$ 
to $\bbZ{2M}$. In fact, at low energies we expect the appearance of a gluino condensate which breaks this further 
to \bbZ{2}. In supergravity this is reflected in the fact that the solution (\ref{eq:KTmetric}) is singular 
at $r=r_0$. The appropriate, non singular metric is the Klebanov Strassler (KS) background \cite{Klebanov:2000hb} where from the
original $U(1)$ isometry only a \bbZ{2} subgroup remains. This happens because for $ds_{[6]}^2$ one uses a metric on the 
deformed conifold $w_1^2+w_2^2+w_3^2+w_4^2=\epsilon^2$, $\epsilon\neq 0$. The 10-dimensional KS metric can be written 
as \cite{Klebanov:2000hb}
\begin{eqnarray}
ds^2 &=& h(\Tau)^{-\half} dx_{\mu}dx^{\mu} + h(\Tau)^{\half} ds^2_{[6]} \\
     &=& h(\Tau)^{-\half} dx_{\mu}dx^{\mu} + h(\Tau)^{\half} 
          \left[\frac{\epsilon^{\frac{4}{3}}}{6}\frac{1}{K(\Tau)^2}d\Tau^2\right]+ ds_{[5]}^2 ,
\label{eq:KSmetric}
\end{eqnarray} 
where we denoted the new radial coordinate as $\Tau$. Also, $ds^2_{[6]}$ is now a Ricci flat, K\"ahler metric on the deformed 
conifold, $ds^2_{[5]}$ is a (\Tau-dependent) metric on a deformed $\Too$. The functions $K(\Tau)$ and $h(\Tau)$ can be obtained 
by solving the equations of motion but will not be of interest here since we will need only the limit $\Tau\rightarrow \infty$ 
where this metric reduces to (\ref{eq:KTmetric}).  

Going back to  the metric (\ref{eq:KTmetric}) we can observe that it is similar to $\adss{5}{5}$ 
but with $S^5$ replaced by $\Too$. Also, in \ads{5} we have $h(r)=R^4/r^4$. Comparing with (\ref{eq:h(r)}) we see that we can 
pictorially describe the background as \ads{5} with a (logarithmically) $r$ dependent radius $R(r)$.  
 One can also define a radial dependent $N_{\mathrm{eff.}}(r)$ by integrating the $5$-form flux over \Too\
and obtaining    
\begin{equation}
N_{\mathrm{eff.}} = N + \frac{3}{2\pi} g_s M^2 \ln\frac{r}{r_0} .
\label{Neff}
\end{equation}
Since we know that \adss{5}{5} with $N$ units of 5-form flux describe a theory with $N^2$ degrees of freedom,
these qualitative considerations lead one to consider that the effective number of degrees 
of freedom changes with the radius, and therefore with the scale. This idea has actually been 
supported by thermodynamic arguments \cite{Buchel:2000ch,Buchel:2001gw,Gubser:2001ri} and from the 
analysis of $2$-point functions in this 
background \cite{Krasnitz:2002ct,Krasnitz:2000ir}. 
 We should point out that since the KT background is singular at $r=r_0$, the KS (which resolves the singularity) 
is necessary to understand the infrared properties of the theory but,  since we are interested in the UV properties of the theory,
which correspond to $r\rightarrow\infty$ and both backgrounds behave similarly in that region,  
we can use the simpler background (\ref{eq:KTmetric}).

\subsection{\label{cascadingft} Cascading field theory}

 The dual field theory to (\ref{eq:KTmetric}) was suggested in \cite{Klebanov:2000nc} and made precise in \cite{Klebanov:2000hb}. 
It was argued that the dual field theory can be described by a set of theories, each valid at a 
certain energy scale, which are \N{1} gauge theories with gauge group $SU(N)\times SU(N+M)$. 
Each theory is given by a different $N(\mu)$ and is valid at a certain energy scale $\mu$.

\begin{figure}
\centerline{\epsfxsize=15cm\epsfbox{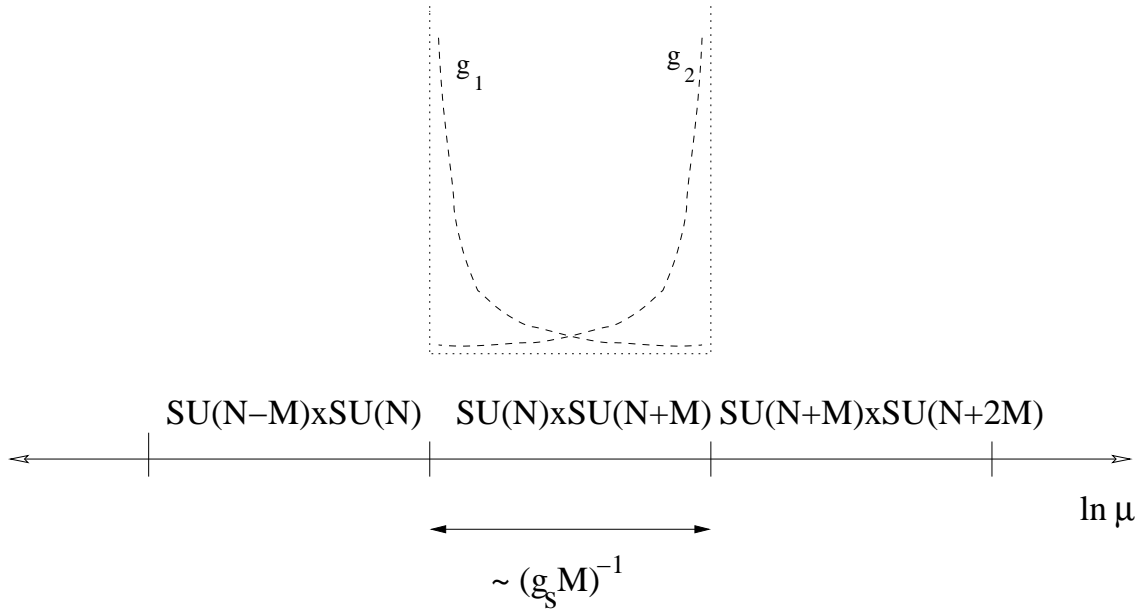}}
\caption{\small Cascading theory.}
\label{fig:cascade}
\end{figure}

 Each theory has the field content of pure \N{1} SYM with gauge group $SU(N)\times SU(N+M)$ plus 
a pair of chiral multiplets $A_{i=1,2}$ in the $(N,\overline{N+M})$ and another 
pair $B_{i=1,2}$ in the $(\bar{N},N+M)$. The global symmetry is $SU_A(2)\times SU_B(2)\times U_B(1)$ where
the $SU_A(2)$ rotates the $A_{i=1,2}$ fields and $SU(2)_B$ rotates the $B$'s. The coupling constants 
 of the two gauge groups run as \cite{Klebanov:2000nc,Klebanov:2000hb}
\begin{eqnarray}
\frac{8\pi^2}{g^2_{N+M}} &=& 3M \ln\left(\frac{\mu}{\mu_1}\right) +\cO(\frac{M^2}{N^2}), \ \ \ \ \mu>\mu_1 , \\
\frac{8\pi^2}{g^2_{N}} &=& 3M \ln\left(\frac{\mu_2}{\mu}\right) +\cO(\frac{M^2}{N^2}), \ \ \ \ \mu<\mu_2  .\\
\label{betafunctions}
\end{eqnarray}
We see that the $SU(N+M)$ gauge theory is asymptotically free whereas the $SU(N)$ is IR free.
The perturbative description is valid only for scales $\mu$ such that $\mu_1<\mu<\mu_2$. For 
$\mu<\mu_1$ the $SU(N+M)$ gauge theory becomes strongly coupled and its correct description 
is in terms of a Seiberg dual theory \cite{seiberg} with gauge group $SU(N-M)$, \ie the gauge group is now
$SU(N-M)\times SU(N)$. For $\mu>\mu_2$ the gauge group $SU(N)$ is dualized to $SU(N+2M)$. It is
interesting that a careful application of Seiberg duality shows that the dual theories have
similar field content. We obtain in this way a cascade of theories which are the same except
for the change in the gauge group (we summarize the situation in fig.\ref{fig:cascade}). It is
meaningful then to define a scale dependent $N$. The 
dependence is such that $N$ changes by $M$ in the interval $\mu_1<\mu<\mu_2$. We get then
\begin{equation}
\frac{\Delta N}{\Delta \ln\mu} = \frac{M}{\ln\frac{\mu_2}{\mu_1}} .
\label{DeltaNmu}
\end{equation}
Using (\ref{betafunctions}) and the supergravity relation \cite{Klebanov:2000hb} 
\begin{equation}
\frac{8\pi^2}{g_{N+M}^2}+\frac{8\pi^2}{g_{N}^2} = \frac{2\pi}{g_s} ,
\end{equation}
we get
\begin{equation}
\frac{8\pi^2}{g_{N+M}^2}+\frac{8\pi^2}{g_{N}^2} = 3M\ln\frac{\mu_2}{\mu_1} = \frac{2\pi}{g_s} \ \ \Rightarrow \ \ 
\ln\frac{\mu_2}{\mu_1} = \frac{\pi}{3Mg_s} ,
\end{equation}
and therefore eq.(\ref{DeltaNmu}) becomes
\begin{equation}
\frac{\Delta N}{\Delta \ln\mu} = \frac{M}{\ln\frac{\mu_2}{\mu_1}} = \frac{3M^2g_s}{2\pi} .
\end{equation}
In the limit where $g_sM^2\ll N$ it is appropriate to define a continuously varying $N$ which can be obtained
by integrating the previous relation:
\begin{equation}
N(\mu) = N_0 + \frac{3}{2\pi} g_s M^2 \ln\frac{\mu}{\mu_0} ,
\end{equation}
in exact agreement with the supergravity result (\ref{Neff}). Furthermore since we are describing the theory
with a continuously varying $N$ it only makes sense to study operators which are, in a certain sense, self dual
under the Seiberg dualities. We can call them operators that cascade nicely. One such type of operator is
\begin{equation}
\tr\left(A_i(x) B_j(x)\right) .
\end{equation}
This operator becomes a meson operator under Seiberg duality. In this case however \cite{Klebanov:2000hb}, meson 
operators are massive and can be integrated out with the result that the operator dualizes (up to factors) into
\begin{equation}
\tr\left(a_i(x) b_j(x)\right) ,
\end{equation}
where $a_i$ and $b_j$ are the chiral superfields of the dual theory. In this sense the operator cascades nicely.


\section{Light-like wedges in the KT-background}
\label{sec:Wl}

\begin{figure}
\centerline{\epsfysize=10cm\epsfbox{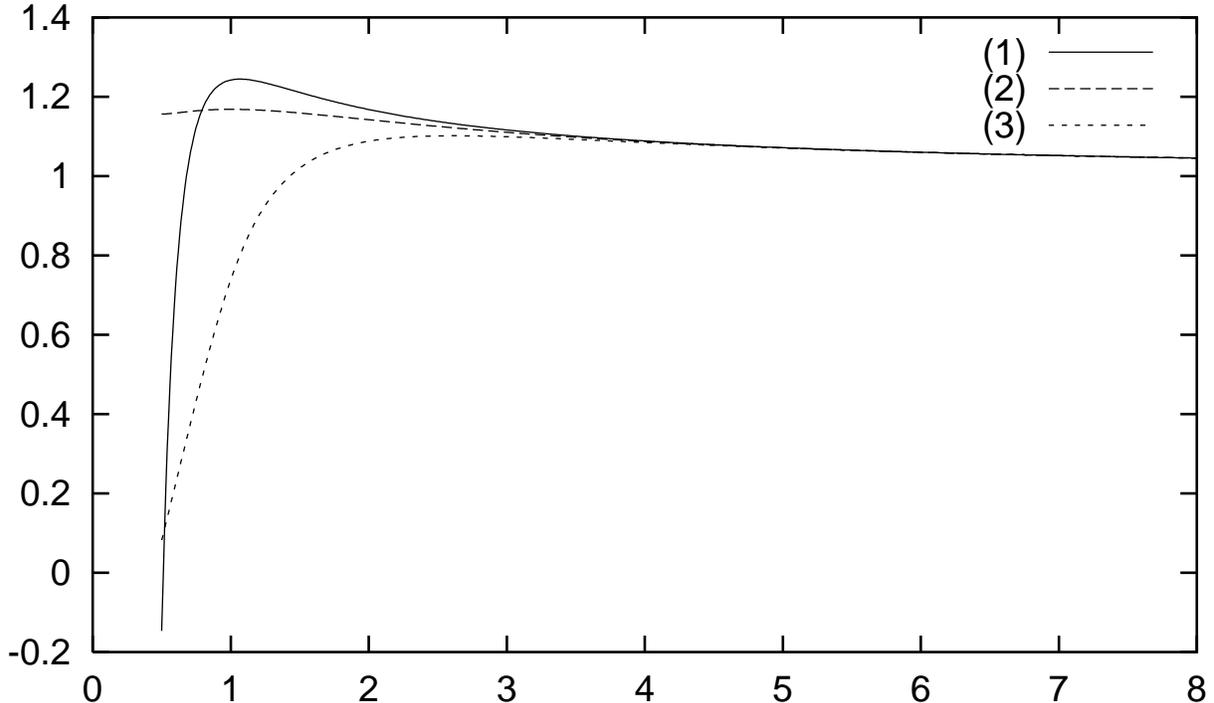}}
\caption{\small The solid line $(1)$ depicts the function (\ref{yexp}) after changing variables to $\zeta=1/x$. The other two are 
numerical solution to eq.(\ref{yeq}) with initial conditions ($y(1)=1.16837534663212$, $y'(1)=0$) for curve $(2)$ 
and ($y(1)=0.7381269451991$, $y'(1)=1$) for curve $(3)$. Again we are using $\zeta=1/x$ as the independent variable. The 
initial conditions are chosen such that both functions are finite at $\z\rightarrow\infty$ in which case
their asymptotic behavior is seen to be well described by the solid line, namely eq.(\ref{yexp}).}
\label{fig:numsol}
\end{figure}

 As we explained in section \ref{adimfWl}, the anomalous dimension of twist two operators can be computed by finding a
surface in the bulk which ends on a light-like wedge in the boundary. In order to do this it
is convenient to write the metric (\ref{eq:KTmetric}) using Rindler coordinates ($\rho,\xi$) in the plane ($t,x_3$):
\begin{equation}
ds^2=h(r)^{-\half} \left(-d\rho^2+\rho^2d\xi^2 +dx_1^2+dx_2^2\right) + h(r)^{\half}\left(dr^2+r^2ds^2_{\Too}\right) .
\end{equation}
 The surface in question is given by a function $r(\rho,\xi)$ which minimizes the world-sheet action. 
Lorentz invariance however implies that $r$ is a function only of $\rho$: 
\begin{equation}
r(\rho,\xi) \stackrel{\mathrm{Lorenz\ inv.}}{\longrightarrow} r(\rho)  .
\end{equation}
 The action that we should minimize becomes then
\begin{equation}
S = \frac{1}{2\pi\alp}\int d\xi\int \rho d\rho \sqrt{r'^2-h^{-1}} = \frac{1}{2\pi\alp} \int d\xi \int \rho(r) \sqrt{1-\rho'^2/h(r)} dr ,
\end{equation}
where we find convenient to solve for $\rho(r)$ instead of $r(\rho)$. At this point we should replace $h(r)$ for its value
from (\ref{eq:h(r)}):
\begin{equation}
h(r) = \frac{1}{R^4} \left[R^4+\half L^4+2L^4\ln\frac{r}{r_0}\right] = \frac{2L^4}{r^4}\ln\frac{r}{\trz} ,
\end{equation}
where, for concision, we introduced $\trz$ defined through 
\begin{equation}
\ln\frac{r_0}{\trz}=\frac{R^4}{2L^4}+\frac{1}{4} .
\end{equation}
 It turns out to be useful to introduce also the definitions
\begin{equation}
r = \trz e^\zeta,\ \ \ \ \trho=\frac{\trz}{\sqrt{2}L^2} \rho .
\end{equation}
The action becomes
\begin{equation}
S = \frac{\sqrt{2}L^2}{2\pi\alpha'} 
    \int d\xi\int  d\zeta e^\zeta \trho \sqrt{1-\frac{1}{\zeta}e^{2\zeta}\left(\frac{d\trho}{d\zeta}\right)^2}  .
\end{equation}
From here we obtain the equation of motion:
\begin{equation}
-\frac{d}{d\zeta}\left[  \frac{\trho e^{3\zeta}}{\zeta\sqrt{1-\frac{1}{\zeta}e^{2\zeta}\left(\frac{d\trho}{d\zeta}\right)^2} }
\left(\frac{d\trho}{d\zeta}\right)\right] = e^\zeta \sqrt{1-\frac{1}{\zeta}e^{2\zeta}\left(\frac{d\trho}{d\zeta}\right)^2}  .
\label{eqmotion}
\end{equation}
 This equation cannot be solved analytically but we are interested actually in the behavior for $r\rightarrow\infty$, 
\ie $\zeta\rightarrow\infty$ because that is the UV region which determines the anomalous dimensions. In that region
we will be able to obtain an asymptotic series expansion of the solution which can be complemented by a numerical analysis.
Before that however let us analyze the properties of the equation. To do that it is useful to introduce a new variable $x$
and function $y(x)$ through
\begin{equation}
\trho = e^{-\zeta} \sqrt{\zeta} \frac{1}{\sqrt{2}} y(x),\ \ \ \ \  x = \frac{1}{\zeta}  .
\label{trho}
\end{equation}
 The resulting equation for $y(x)$ can be written as
\begin{equation}
y''(x) = \frac{1}{y x^4} P(x,y,y') ,
\label{yeq}
\end{equation}
where $P(x,y,y')$ is a polynomial given by
\begin{eqnarray}
P(x,y,y') &=& - (y^2-1)(y^2-2) +\left (-\frac{5}{2}+\frac{7}{4} y^2\right )y^2 x+
\left (3  y'-\frac{9}{8}y^3-3y^2y'+y\right )y x^2 \nonumber \\ 
&& +\left (\frac {15}{4}y^2y'-3y'+\frac{5}{16}y^3\right )y x^3
+\left (-\frac{3}{2}\,{y}^{3}{  y'}-3\,{y}^{2}{{  y'}}^{2}+{{  y'}}^{2}-\frac{1}{32}\,{y}^{4}\right ){x}^{4} \nonumber\\
&&+\left (\frac{9}{4}   y'^2+\frac{3}{16} y  y'\right )y^2x^5
+\left (-  y'^3-\frac{3}{8}yy'^2\right )yx^6
+\frac{1}{4}\,y y'^3x^7 .
\end{eqnarray}
 Although the actual expression for the polynomial is not very illuminating the point is that in the form (\ref{yeq}) 
it is clear that the equation has no singular points except at
$x=0$ and $y=0$. That means that around any value $x_0\neq 0$ we can find a solution with given initial 
arbitrary values $y(x_0)\neq 0$ and $y'(x_0)$. This solution can be found numerically or in a power series 
expansion that converges in a neighborhood of $x_0$. On the other hand we are interested in the region 
$\zeta\rightarrow\infty$ that corresponds to $x\rightarrow 0$, namely the behavior near the singularity.
If we expand $P(x,y,y')$ for small $x$ we get
\begin{equation}
P(x,y,y') = - (y^2-1)(y^2-2) + \cO(x) ,
\end{equation}
which means that if we want $y''(x)$ (and $y(x)$) to stay finite as $x\rightarrow 0$, we need 
(from eq.(\ref{yeq})) that $y(0)=\pm 1$ or $y(0)=\pm\sqrt{2}$.
We will see later on that $y(0)=\pm\sqrt{2}$ corresponds to a light-like surface whereas $y(0) =\pm 1$ corresponds
to the Euclidean surface we are looking for. In the \N{4} case in Poincare coordinates, this correspond to having the 
surfaces $z^2=t^2-x^2$ and $z^2=2(t^2-x^2)$, the latter being the relevant one. After we choose $y(0)=1$ the next
step is to observe that $y'(0)$ is also determined if the solution is not to be singular and so on. In that way we
get the asymptotic expansion 
\begin{equation}
y(x) = \left[1+\frac{3}{8}\, x - \frac{13}{128}\, x^2 
       + \frac{127}{1024}\, x^3 - \frac{5085}{32768} x^4 +\ldots \right] + \cO\left(e^{-\sqrt{2}/x}\right) .
\label{yexp}
\end{equation}
Within the square parenthesis we have an infinite series all whose coefficients are completely determined. To this,
an arbitrary function of order $\cO\left(e^{-\sqrt{2}/x}\right)$ can be added. This function has all its derivatives equal to zero and
cannot be derived from the condition of finiteness at $x=0$. It is determined by the initial conditions. That this has to be the case 
is clear from the previous discussion. For each initial point $x_0$ we have a two parameter family of solutions parameterized 
by $y(x_0)$ and $y'(x_0)$. The condition of finiteness at $x=0$ gives a relation between $y(x_0)$ and $y'(x_0)$ but there is still a one 
parameter family of solutions regular at $x=0$. What we stated above is that all those solutions have the asymptotic 
behavior (\ref{yexp}) up to exponentially suppressed terms. We can verify this numerically by plotting different solutions and 
comparing with eq.(\ref{yexp}). We do that in fig.\ref{fig:numsol} where we plotted $y(\zeta)$ for $y(1)=1.16837534663212$, $y'(1)=0$ 
and $y(1)=0.7381269451991$, $y'(1)=1$ together with the function (\ref{yexp}) after changing variables $x=1/\zeta$. We see that the 
three function have the same asymptotic behavior, the difference can also be computed and agrees with the analytical result that it 
is exponentially suppressed. This estimate follows from expanding the polynomial $P$ around $x=0$, $y=1$ which gives
\begin{equation}
y \simeq 1 +\delta y, \,\,\,\mbox{with}\ \ \ \ x^4\,\delta y''-2\, \delta y = -\frac{3}{4} x .
\end{equation}
The linear equation for $\delta y$ (together with the condition of $\delta y$ finite at $x=0$) can be exactly solved: 
\begin{equation}
\delta y = \frac{3}{8} x + C_1 x e^{-\sqrt{2}/x} .
\end{equation}
From here we get the value $y'(0)=3/8$. The constant $C_1$ is arbitrary since it multiplies a solution of the homogeneous equation and
corresponds to the arbitrary function that we mentioned before has to be determined by the initial conditions. This initial 
conditions should follow from the IR of the background but as we see are not relevant as long as we discard the exponentially 
suppressed terms. 
 
Going back to eq.(\ref{eqmotion}) we find, from (\ref{yexp}) and (\ref{trho}) that the asymptotic solution is 
\begin{equation}
\trho(\z) \sim e^{-\z} \sqrt{\z} \frac{1}{\sqrt{2}} \left[1+\frac{3}{8}\, \frac{1}{\z} - \frac{13}{128}\, \frac{1}{\z^2} 
                                   + \frac{127}{1024}\, \frac{1}{\z^3} - \frac{5085}{32768} \frac{1}{\z^4} + \ldots\right] .
\end{equation}
The previous discussion was mainly to understand the properties of the equation and what the asymptotic expansion means. For
the purpose of computing the anomalous dimension we will keep only the first term which gives, after replacing in the action
\begin{equation}
\left.\begin{array}{lcl}
\trho &\simeq& e^{-\zeta} \frac{\sqrt{\zeta}}{\sqrt{2}} \\
\frac{d\trho}{d\zeta} &\simeq& -e^{-\zeta}\frac{\sqrt{\zeta}}{\sqrt{2}} 
\end{array} \right\}\ \  
\Rightarrow \ \ S\simeq \frac{\sqrt{2}L^2}{2\pi\alp} \int d\xi\int d\zeta \frac{\sqrt{\zeta}}{\sqrt{2}} \sqrt{1-\half} .
\end{equation}
If we had chosen $y(0)=\sqrt{2}$, inside the square root we would have got $\sqrt{1-1}=0$, namely a light-like surface as 
we said before. We can regulate the integrals in the action in the same manner as for the \N{4} case, integrating the angle
$\xi$ between $-\gamma/2$ and $\gamma/2$ with $\gamma\rightarrow\infty$ and $\zeta$ between $0$ and $\zeta_m$:
\begin{equation}
S = \frac{L^2}{2\pi\alp\sqrt{2}} \int_{-\gamma/2}^{\gamma/2} d\xi\int_0^{\zeta_m} d\zeta \sqrt{\zeta} = 
    \frac{L^2}{2\pi\alp\sqrt{2}} \gamma \frac{2}{3}\zeta_m^{\frac{3}{2}} .
\end{equation}
 The parameter $\zeta_m$ should be understood as a UV cutoff since $\zeta=\ln(r/\trz)$ and then $\zeta<\zeta_m$ is the same 
as $r<r_m=\trz\exp(\zeta_m)$. In the cascade theory, if the steps of the cascade are large ($g_sM^2\sim N$) then one can introduce
a UV cut-off for each theory (\eg $\mu_2$ in our previous notation) but in the limit ($g_sM^2\ll N$) the steps of the cascade are small
and the cutoff should be better understood as the energy scale $\mu$ at which we probe the cascade which just determines the value
of $N_{\mbox{eff.}}(\mu)$ that we have to use. Then we identify $\zeta_m=\ln(\mu/\mu_0)$ where we identified $r\sim \mu$ up to
a constant. We get then
\begin{equation}
S = \ln W = \frac{L^2}{2\pi\alp\sqrt{2}} \gamma \frac{2}{3}\left(\ln\frac{\mu}{\mu_0}\right)^{\frac{3}{2}} .
\end{equation}
Now we appeal to the definition of anomalous dimension
\begin{equation}
 \mu\frac{\partial W}{\partial\mu} = -\Gc W ,
\label{andimdef}
\end{equation}
and get
\begin{equation}
\Gc= -\frac{L^2}{2\pi\alp\sqrt{2}} \gamma \left(\ln\frac{\mu}{\mu_0}\right)^{\frac{1}{2}} ,
\end{equation}
which is linear in $\gamma$ as expected from Lorentz invariance. The proportionality coefficient is $\bGc$:
\begin{equation}
\bGc = -\frac{L^2}{2\pi\alp\sqrt{2}} \left(\ln\frac{\mu}{\mu_0}\right)^{\frac{1}{2}} ,
\end{equation}
from where the Korchemsky-Marchesini formula (\ref{eq:adfWl}) gives the anomalous dimension
\begin{equation}
\gamma(\mu) = 2 \frac{L^2}{\pi\alp\sqrt{2}} \left(\ln\frac{\mu}{\mu_0}\right)^{\frac{1}{2}} \ln S 
            = \frac{3\sqrt{3}}{2}\frac{1}{\sqrt{\pi}} \sqrt{g_sN_{\mbox{eff.}}} \ln S ,
\label{Wlresult}
\end{equation}
where the extra factor of two appears because the Wilson loop should actually be in the adjoint (twice the fundamental). 
In the second equality we used the definition (\ref{eq:Ldef}). This anomalous dimension is the main result of this section 
and of the paper. In the next section we reproduce it using rotating strings. The important difference with the \N{4} case 
is that the theory is not conformally invariant and so the anomalous dimension depends on the scale and we have to use the 
appropriate definition of anomalous dimension as eigenvalues under infinitesimal scale transformations (\ref{andimdef}).
Incidentally taking the derivative (\ref{andimdef}) provides a factor $3/2$ that makes the result agree with the rotating 
string calculation.


\section{Rotating string version}
\label{sec:KTrotstring}

 Here we do the rotating string calculation equivalent to the Wilson loop calculation of the previous 
section\footnote{There is a large literature on rotating strings in different backgrounds. For the case 
in hand the closest related work is the recent paper \cite{Schvellinger:2003vz} and 
also \cite{Tseytlin:2002ny} where some of the results of this section where anticipated.}. For that we will first 
consider the KS background (\ref{eq:KSmetric})
\begin{equation}
ds^2 = h^{-\half}(\Tau) (-dt^2+dx_1^2+dx_2^2+dx_3^3) + h^{\half}(\Tau) \frac{ \varepsilon^{\frac{4}{3}} }{6} 
\frac{1}{K^2(\Tau)} d\Tau^2 + ds_5^2(\Tau) .
\end{equation}
We can do a coordinate transformation
\begin{equation}
z = \int_\Tau^\infty \frac{\varepsilon^{\frac{4}{3}}}{6} 
\frac{h(\Tau)}{K^2(\Tau)} ,
\label{zdef}
\end{equation}
to render the metric into the form
\begin{equation}
ds^2 =  h^{-\half}(\Tau(z)) (dz^2-dt^2+dx_1^2+dx_2^2+dx_3^3) + ds_5^2(\Tau(z)) .
\end{equation}
Now the same coordinate transformation as in (\ref{Poinnew}) results in the metric:
\begin{eqnarray}
ds^2 &=&  h^{-\half}(\Tau(\frac{e^{-\tau}}{\cosh(\rho)})) \frac{e^{-2\tau}}{\cosh^2\rho} 
        \left[ \cosh^2\rho d\tau^2 + d\rho^2 + \sinh^2\rho d\phi^2 + \sinh^2\rho\sin^2\phi d\theta^2 \right.\\
     & &\left.  -\sinh^2\rho\cos^2\phi d\chi^2\right] + ds_5^2(\tau,\rho)  .
\end{eqnarray}
Naturally it is difficult to find the equivalent solution to that of eq.(\ref{rotansatzN4}) in this background. However
since we are interested in the asymptotic properties of the string in the UV, this corresponds to taking
$\Tau\rightarrow\infty$ and we can use the Klebanov-Tseytlin solution. Using a new radial coordinate 
$r=3^{\half} 2^{-\frac{5}{6}} \epsilon^{\frac{2}{3}} e^{\frac{\tau}{3}}$ and after an appropriate identification
of the constants, we recover the metric (\ref{eq:KTmetric}). The transformation (\ref{zdef}) becomes
\begin{equation}
z = \int^r h(r)^{\half} dr \simeq \frac{\sqrt{2}L^2}{r} \left(\ln\frac{r}{\trz}\right)^{\half},\ \ \ \ (r\rightarrow\infty),
\end{equation}
where the last equality is valid for large $r$ and we used that, as before,  $h(r)$ is given by
\begin{equation}
h(r) = \frac{2L^4}{r^4} \ln\frac{r}{\trz} .
\end{equation}
Note that since $\cosh\rho>1$ and $z=e^{-\tau}/\cosh\rho$ the string never extends beyond $z=e^{-\tau}$, namely, we do not need to 
know the IR part of the background if $\tau\gg1$.  
In this limit, and taking into account that the string does not extend into $x_{1,2}$ or the internal coordinates 
we can write the relevant part of the metric as
\begin{equation}
ds^2 \simeq z^2h^{-\half} \left(\cosh^2\rho d\tau^2 + d\rho^2 -\sinh^2\rho d\chi^2\right) .
\end{equation}
Now we compute, at leading order for large $r$ (or large $\tau$):
\begin{eqnarray}
z^2h^{-\half} &\simeq& \sqrt{2}L^2 \left(\ln\frac{r}{\trz}\right)^{\half} 
              \simeq \sqrt{2}L^2 \left(-\ln z\right)^{\half}  \\
              &=& \sqrt{2}L^2 \left(\ln(e^{-\tau}/\cosh\rho)\right)^{\half}
              = \sqrt{2}L^2 \left(\tau+\ln\cosh\rho\right) .
\end{eqnarray}
Within this approximation the metric becomes
\begin{equation}
ds^2 =   \sqrt{2}L^2 \sqrt{\ln(\cosh\rho) +\tau} \left(\cosh^2\rho d\tau^2 + d\rho^2 -\sinh^2\rho d\chi^2\right) .
\end{equation}
As a further approximation we can use that for $\tau\rightarrow\infty$, we have $\tau \gg \ln(\cosh\rho)$ since
$\rho<\rho_0$ and $\rho_0$ is fixed. 
So we finally get
\begin{equation}
ds^2 = \sqrt{2}L^2 \sqrt{\tau} \left(\cosh^2\rho d\tau^2 + d\rho^2 -\sinh^2\rho d\chi^2\right) .
\end{equation}
As a last approximation we note that $\frac{d}{d\tau} \sqrt{\tau} = 1/\sqrt{\tau} \rightarrow 0$ and then
an adiabatic approximation where $\tau$ is considered constant is appropriate (for large $\tau$). In this case
we can define a $\tau$-dependent momentum $P_\tau$ which is the bulk equivalent of the fact that the anomalous 
dimension runs with the scale. Namely, the theory is not scale invariant, but we can define for each scale 
a conformal dimension that contains the information of how an operator changes under an infinitesimal scale
transformation. This conformal dimension is scale dependent because $P_\tau$ is $\tau$-dependent.

 If we consider $\tau$ constant the calculation becomes the same calculation that we did in section \ref{adimfrt}
but after replacing $R^2\rightarrow \sqrt{2}L^2 \sqrt{\tau}$ in the metric (\ref{N4metric}). This gives (\ref{N4rotstring})
with the same replacement $R^2\rightarrow \sqrt{2}L^2 \sqrt{\tau}$:
\begin{equation}
\Delta \simeq S +  \frac{\sqrt{2}L^2}{\pi\alp} \sqrt{\tau} \ln S , \ \ \ \ \ (S\rightarrow\infty) .
\label{rotstringN1}
\end{equation}
Since at leading order we have $\ln r = \tau$ and we identify $r$ with the scale $\mu$ the result reads
\begin{equation}
\Delta = \frac{2 L^2}{\pi\alp} \left(\ln\frac{\mu}{\mu_0}\right)^{\frac{1}{2}} \ln S 
       = \frac{3\sqrt{3}}{2}\frac{1}{\sqrt{\pi}} \sqrt{g_sN_{\mbox{eff.}}} \ln S ,
\end{equation}
in exact agreement with the Wilson loop calculation (\ref{Wlresult}).


\section{Field theory interpretation}
\label{sec:FT}

 In the previous two sections we used the method of Wilson loops and the one of rotating strings to compute anomalous dimensions 
of certain field theory operators. In this section we discuss in more detail which are the operators involved. A related
discussion already appeared in \cite{Buchel:2002yq}.  The first point is that since we are in the regime where the steps of 
the cascade are small compared to $N$, the cascade should be considered as a continuum of theories. In that case the only operators 
that make sense are those that cascade nicely, that is that can be defined at each step of the cascade and preserve 
their form under the Seiberg dualities.

In section \ref{sec:KTbkg} we studied the operators 
\begin{equation}
\cO_{ij} = \tr\left(A_i(x) B_j(x)\right) .
\end{equation}
They can be generalized to bilocal operators:
\begin{equation}
\cO_{ij}(\Delta_\mu) = \tr\left(A_i(x+\Delta) e^{\int_{x}^{x+\Delta}(A^{(M+N)}_\mu+A^{(N)}_\mu) dx^\mu} B_j(x)\right) ,
\label{Oij}
\end{equation}
where we inserted a Wilson line in the bi-fundamental of $SU(N+M)\times SU(N)$ to preserve gauge invariance. These
operators should cascade to 
\begin{equation}
\cO_{ij}(\Delta_\mu) = \tr\left(a_i(x+\Delta) e^{\int_{x}^{x+\Delta}(A^{(M+N)}_\mu+A^{(N)}_\mu) dx^\mu} b_j(x)\right) ,
\end{equation}
since they are bilocal gauge invariant operators with the same transformation properties under
the global symmetry $SU_A(2)\times SU_B(2)\times U_B(1)$. The actual operators that cascade nicely
can differ from these ones by functions of the fundamental fields that are invariant under the
global symmetries (see \cite{Klebanov:2000hb} for a related discussion). We just consider this
discussion as evidence of the existence of such operators and use (\ref{Oij}) as a guide to understand their
properties. 

 If we expand (\ref{Oij}) in powers of $\Delta_\mu$ we get a series of operators
\begin{eqnarray}
\cO_{ij}(\Delta_\mu) &=& \tr\left(A_i(x+\Delta) e^{\int_{x}^{x+\Delta}(A^{(M+N)}_\mu+A^{(N)}_\mu) dx^\mu} B_j(x)\right) 
= \sum_{S=0}^{\infty} \frac{1}{S!} \cO_S^{ij}(\Delta_\mu) ,\\
\cO_S^{ij}(\Delta_\mu) &=& \tr\left( A_i \nabla_{\mu_1} \ldots \nabla_{\mu_S} B_j\right) \Delta^{\mu_1} \ldots \Delta^{\mu_S} .
\end{eqnarray}
 If we take, as we did before,  $\Delta_\mu$ to be light-like then the operators that survive in the contraction are those
of maximum helicity, \ie\ a certain component of the operators of maximum spin that we can form for a given 
number of derivatives. Since the number of derivatives determines the conformal dimension (in the free theory) these 
are the operators of lower twist. Note that in that case the Wilson line that appears in eq. (\ref{Oij}) is  light-like. 
 If one tries to compute such a Wilson line using the AdS/CFT correspondence the natural thing is to use a world-sheet that ends
on such a line at the boundary. It is interesting to note that the world-sheet we considered, which appeared as a double Wick 
rotation of the rotating string has precisely such property. It might be interesting to analyze this point further since a 
light-like Wilson line is BPS (since its variation involves $\Gamma^\mu \Delta_\mu$ which 
satisfies $(\Gamma^\mu \Delta_\mu)^2=\Delta^2=0$). 

 Going back to our main point, it is reasonable to suggest that the operators whose anomalous dimension we are computing 
are precisely $\cO_S^{ij}$. However, we finalize by pointing out that one can use the vector multiplets to construct other 
operators which seem to cascade nicely and are invariant under the global symmetries. This is because under Seiberg 
duality the field strength $W_{\alpha}^2$  dualizes to $-\tilde{W}_{\alpha}^2$. This operators should
be also taken into account if a precise matching between supergravity and the field theory is required. 
However the supergravity calculation is presently not able to distinguish between all the possible 
twist two operators. The same happens in the \N{4} case where there are several twist two operators (\ref{eq:t2op}).
 The way to distinguish between them is their transformation properties under the global symmetries
and supersymmetry. The natural conjecture is that all operators that one can form have the same anomalous dimension
for large spin. 


\section{Conclusions}  
\label{sec:conclusions}

In this paper we extended previous calculations of anomalous dimensions of twist two operators
that were done for \N{4} SYM theory to the cascading theories which have \N{1} supersymmetry.
 The interesting difference is that in the \N{1} case the anomalous dimension depends on the scale
since the theory is not conformally invariant. To tackle the non-conformal case we needed to use
a variation of the rotating string method which makes explicit use of the fact that the anomalous dimension 
is a UV property and then only the UV region of the background should be involved in the calculation. 
Furthermore we did not use global coordinates since in the non-conformal case there is no state-operator 
map. In general, a similar calculation done in coordinates akin to global coordinates 
computes the energy of a certain high spin `glueball' state which is not directly related
to the anomalous dimensions of an operator unless the theory is conformally invariant. 

 Our main interest was in the Wilson loop method that can be easily applied in this case.
It makes evident that the logarithmic behavior of the anomalous dimension is a consequence of Lorentz 
invariance and so seems to be a generic property of all theories that have supergravity duals. In 
field theory it is known that those are gauge theories since in scalar theories twist two operators 
have vanishing anomalous dimension in the limit of large spin. 

 In view of the fact that certain Wilson loops \cite{Erickson:2000af} can be computed at strong coupling by summing over
a subset of Feynman diagrams we believe to be of high interest to see if a similar calculation 
can give the anomalous dimensions that are obtained from supergravity. 


\begin{acknowledgments}
I am very grateful to I. Klebanov, J. Maldacena, R. Myers and M. Strassler for comments and
discussions. I also want to thank C. Herzog and A. Tseytlin  for some comments on a previous
version of this paper. This work was supported in part by NSF through grant PHY-0331516. 
In addition I would like to thank the Perimeter Institute for Theoretical Physics 
for hospitality and partial support while this work was being done. 
\end{acknowledgments}


\appendix*

\section{Coordinates in \ads{5}}

 $AdS_5$ has an $SO(4,2)$ isometry group. The isometries are obviously independent of the 
system of coordinates used, but different systems of coordinates make different isometries 
manifest (meaning that the coordinates transform simply under a subgroup of the full 
isometry group) and therefore for different purposes different coordinates are useful. 

First consider embedding coordinates. Introducing Cartesian coordinates ($X$, $Y$, $Z$, $W$, $U$, $V$) 
in $\bbR{(4,2)}$, \ads{5} is the manifold defined by the equation:
\begin{equation}
-X^2-Y^2-Z^2-W^2+U^2+V^2 = R^2,
\end{equation}
for some arbitrary constant $R$. The metric is the one induced by the one in $\bbR{(4,2)}$:
\begin{equation}
ds^2 = dX^2 + dY^2 + dZ^2 + dW^2 -dU^2 - dV^2.
\end{equation}
 In this (redundant) coordinates, the $SO(4,2)$ isometry group is manifest. More close to the
field theory are the Poincare coordinates which parameterize the patch $U+W>0$ and are given by:
\begin{equation}
\begin{array}{lll}
X = R\, {x_1/ z}, &  Y = R\, {x_2/ z}, &  Z = R\, {x_3/ z}, \\
W = -{1\over 2z}(-R^2+z^2+x_i^2-t^2), & U = {1\over 2z}(R^2+z^2+x_i^2-t^2), & V=R\,{t/ z},
\end{array}
\end{equation}  
with a metric
\begin{equation}
ds^2 = \frac{R^2}{z^2}\left(dz^2+dx_i^2-dt^2\right) .
\end{equation}
 In this coordinates we have a manifest $SO(3,1)\times SO(1,1)\subset SO(4,2)$ corresponding 
to Lorentz transformations and scale transformations $z\rightarrow \lambda z$, $x_i\rightarrow \lambda x_i$,
$t\rightarrow \lambda t$. Scale transformations correspond to boosts along $U,W$ and that is why we denote them
as an $SO(1,1)$ subgroup. Operators in the field theory are classified by their spin $(j_1,j_2)$ and their
conformal dimension $\Delta$.  

Global coordinates on the other hand are introduced by splitting the Cartesian coordinates in two groups
satisfying:
\begin{equation}
U^2+V^2 = R^2 \cosh^2\rho, \ \ \ \ \ X^2+Y^2+Z^2+W^2 = R^2 \sinh^2\rho .
\end{equation}
For fix $\rho$, the coordinates ($X$,$Y$,$Z$,$W$) span a sphere $S^3$ and the coordinates $(U,V)$ a sphere
$S^1$ which we parameterize with an angle $t$. The metric can then be written as
\begin{equation}
ds^2 = R^2(-\cosh^2\!\rho\,dt^2 + d\rho^2 + \sinh^2\!\rho\,d\Omega_{[3]}^2),
\end{equation}
where $\Omega_{[3]}$ is the metric of a unit $3$-sphere.
They make manifest an $SO(4)\times SO(2)\subset SO(4,2)$ symmetry corresponding to rotations 
in the $(X,Y,Z,V)$ and $(U,V)$ spaces. These coordinates are convenient because they cover the whole
of $AdS_5$ space. To relate the eigenvalues under $SO(4)\times SO(2)$ to those of $SO(3,1)\times SO(1,1)$
which are the relevant ones in the field theory we need to do a double Wick rotation such as
$V\rightarrow iV$, $W\rightarrow iW$. Under this transformation, the energy, the eigenvalue under
$SO(2)$ becomes the conformal dimension, the eigenvalue under $SO(1,1)$. 

 In this paper we find convenient to introduce another parameterization where the symmetry $SO(3,1)\times SO(1,1)$
is manifest and rotations under $SO(1,1)$ become translations, namely a conserved momentum. In that
way no Wick rotation is necessary.
 To do that we split the embedding coordinates into two groups which are now
$(U,W)$ and $(X,Y,Z,V)$ and parameterize this as 
\begin{equation}
U^2-W^2 = R^2\cosh^2\rho, \ \ \ \ \ X^2+Y^2+Z^2-V^2=R^2\sinh^2\rho .
\end{equation}
Now, for fixed $\rho$ the coordinates ($X$, $Y$, $Z$, $V$) span a de Sitter space $\Sigma_{[3]}$ and
the coordinates ($U$, $W$) a real line which we parameterize with a coordinate $\tau$. With this splitting
there is an explicit $SO(1,1)$ isometry corresponding to translations in $\tau$ or equivalently to boosts in the plane ($U$,$W$),
and an $SO(3,1)$ isometry which is the isometry group of $\Sigma_{[3]}$. 
The metric is given by 
\begin{equation}
ds^2 = R^2(\cosh^2\rho\, d\tau^2 + d\rho^2 + \sinh^2\rho\, d\Sigma_{[3]}^2), 
\end{equation}
where $d\Sigma^3_{[3]}$ is the  unit metric on the 3-dimensional de Sitter space $\Sigma_{[3]}$. We see that these coordinates
are related to global coordinates by a double Wick rotation: $(t\rightarrow i\tau$, $S^3\rightarrow \Sigma_3$). This
coordinates do not cover the whole space as is evident from the fact that $|U|>|W|$. For the purpose of the paper
it is useful to consider coordinates ($\chi$, $\phi$, $\theta$) on $\Sigma_{[3]}$ which together with $\tau$ and $\rho$ parameterize \ads{5} as: 
\begin{equation}
\begin{array}{lll}
U=R\cosh\rho\cosh\tau, & Z = R\sinh\rho\cos\phi\cosh\chi   & X = R\sinh\rho\sin\phi\cos\theta, \\
W=R\cosh\rho\sinh\tau, & V = R\sinh\rho\cos\phi\sinh\chi,  & Y = R\sinh\rho\sin\phi\sin\theta .
\end{array}
\end{equation}
The metric is 
\begin{equation}
ds^2 = \cosh^2\rho d\tau^2 + d\rho^2 + \sinh^2\rho d\phi^2 + \sinh^2\rho\sin^2\phi d\theta^2-\sinh^2\rho\cos^2\phi d\chi^2 .
\end{equation}
The coordinates chosen cover only part of \ads{5}. In particular we see that $U+W>0$ which implies that
this coordinate patch is completely contained in the Poincare patch. The relation to Poincare coordinates is easy to work out:
\begin{equation}
\begin{array}{lclllcl}
x_1 &=& R\, e^{-\tau}\,\tanh\rho\,  \sin\phi \cos\theta &,&
x_2 &=& R\, e^{-\tau}\,\tanh\rho\,  \sin\phi \sin\theta  , \\
x_3 &=& R\, e^{-\tau}\,\tanh\rho\,  \cos\phi \cosh\chi &,&
t   &=& R\, e^{-\tau}\,\tanh\rho\,  \cos\phi \sinh\chi , \\
z   &=& R\, e^{-\tau}\, \frac{1}{\cosh(\rho)} . &\ \  &&& 
\end{array}
\end{equation}
We can see now that the $SO(1,1)$ isometry which was given by translations in $\tau$ corresponds to 
$z\rightarrow \lambda z$, $x_i \rightarrow \lambda x_i$, namely a scale transformation. So we identify the associated momentum
$P_{\tau}$ with the conformal dimension $\Delta=P_{\tau}$. No Wick rotation is necessary in this case.




\end{document}